\documentclass{jpsj-suppl}
\usepackage{txfonts} 
\usepackage{tikz}
\title{Strangeness Photoproduction at the BGO-OD Experiment}

\author{%
T. C. Jude $^{1}$,
S. Alef$^{1}$,
D. Bayadilov$^{2}$,
R. Beck$^{2}$,
M. Becker$^{2}$,
A. Bella$^{1}$,
P. Bielefeldt$^{1}$,
\mbox{S. Boese$^{2}$},
A. Braghieri$^{3}$,
K. Brinkmann$^{4}$,
P. Cole$^{1}$,
F. Curciarello$^{6,7}$,
V. De Leo$^{6,7}$,
\mbox{R. Di Salvo$^{8}$},
H. Dutz$^{1}$,
D. Elsner$^{1}$,
A. Fantini$^{8,9}$,
O. Freyermuth$^{1}$,
S. Friedrich$^{4}$,
\mbox{F. Frommberger$^{1}$},
V. Ganenko$^{5}$,
G. Gervino$^{10,11}$,
F. Ghio$^{12,13}$,
G. Giardina$^{6,7}$,
S. Goertz$^{1}$,
A. Gridnev$^{15}$,
E. Gutz$^{4}$,
D. Hammann$^{1}$,
J. Hannappel$^{1}$,
P. Hartmann$^{1}$,
W. Hillert$^{1}$,
\mbox{A. Ignatov$^{16}$},
R. Jahn$^{2}$,
R. Joosten$^{2}$,
F. Klein$^{1}$,
K. Koop$^{2}$,
B. Krusche$^{17}$,
A. Lapik$^{16}$,
\mbox{P. Levi Sandri$^{18}$},
I. V. Lopatin$^{15}$,
G. Mandaglio$^{6,7}$,
F. Messi$^{1}$,
R. Messi$^{8,9}$,
V. Metag$^{4}$,
\mbox{D. Moricciani$^{8}$},
A. Mushkarenkov$^{16}$,
M. Nanova$^{4}$,
V. Nedorezov$^{16}$,
D. Novinskiy$^{15}$,
\mbox{P. Pedroni$^{3}$},
B. Reitz$^{1}$,
M. Romaniuk$^{8}$,
T. Rostomyan$^{17}$,
N. Rudnev$^{16}$,
G. Scheluchin$^{1}$,
\mbox{H. Schmieden$^{1}$},
A. Stugelev$^{15}$,
V. Sumachev$^{15}$,
V. Tarakanov$^{15}$,
V. Vegna$^{1}$,
D. Walther$^{2}$,
\mbox{D. Watts$^{14}$},
H. Zaunick$^{2,4}$ and
T. Zimmermann$^{1}$
 }
 \inst{%
 $^{1}$ Physikalisches Institut, Nussallee 12, 53115 Bonn, Germany\\
 $^{2}$ Helmholtz-Institut f\"{u}r Strahlen- und Kernphysik, Nussallee 14-16, 53115 Bonn, Germany\\
 $^{3}$ INFN sezione di Pavia, Via Agostino Bassi, 6 - 27100 Pavia, Italy\\
 $^{4}$ Justus-Liebig-Universit\"{a}t Gie\ss{}en, II. Physikalisches Institut, Heinrich-Buff-Ring 16, 35392 Gie\ss{}en, Germany\\
 $^{5}$ National Science Center Kharkov Institute of Physics and Technology, Akademicheskaya St. 1, Kharkov, 61108, Ukraine\\
 $^{6}$ INFN sezione Catania, 95129 Catania, Italy\\
 $^{7}$ Universit\`{a} degli Studi di Messina, Via Consolato del Mare 41, 98121 Messina, Italy\\
 $^{8}$ INFN Roma Tor Vergata, Via della Ricerca Scientifica 1, 00133 Roma, Italy\\
 $^{9}$ University of Rome ``Tor Vergata'',Physics department, Via della Ricerca Scientifica 1, 00133 Roma, Italy\\
 $^{10}$ INFN sezione di Torino, Via P.Giuria 1, 10125 Torino, Italy\\
 $^{11}$ Dipartimento di Fisica, Universit\`{a} di Torino, via P. Giuria 1, 10125 Torino, Italy\\
 $^{12}$ INFN sezione di Roma, c/o Dipartimento di Fisica - Universit\`{a} degli Studi di Roma ``La Sapienza'' P.le Aldo Moro, 2, 00185 Roma, Italy\\
 $^{13}$ Istituto Superiore di Sanit\`{a}, Viale Regina Elena 299, 00161 Roma, Italy\\
 $^{14}$ The University of Edinburgh, James Clerk Maxwell Building, Mayfield Road, Edinburgh EH9 3JZ, UK\\
 $^{15}$ Petersburg Nuclear Physics Institute, Gatchina, Leningrad District, 188300, Russia\\
 $^{16}$ Russian Academy of Sciences Institute for Nuclear Research, prospekt 60-letiya Oktyabrya 7a, Moscow 117312, Russia\\
 $^{17}$ Institut f\"{u}r Physik, Klingelbergstrasse 82, 4056 Basel, Switzerland\\
 $^{18}$ INFN - LNF, Via E. Fermi 40, 00044 Frascati, Italy\\
 }

\email{jude@physik.uni-bonn.de}
\recdate{June 23, 2011}

\abst{BGO-OD is a newly commissioned experiment
to investigate the internal structure of the nucleon,
using an energy tagged bremsstrahlung photon beam at the ELSA electron facility.  
The setup consists of a highly
 segmented BGO calorimeter surrounding the target, with a particle tracking
 magnetic spectrometer at forward angles. 
BGO-OD is ideal for investigating meson photoproduction.
The extensive physics programme for open strangeness photoproduction is introduced, and preliminary analysis presented.
}

\kword{photoproduction, dynamically generated resonance, strangeness, BGO-OD}

\begin{document}
\maketitle

\section{Introduction}

The excitation spectrum of the nucleon gives an insight to the constituents and their interactions in the non-peturbative QCD range.
Constituent quark models (CQM) inspired from pure colour charge interactions, describe this spectrum with limited success.  For example,  the pattern of the mass and parity of the Roper resonance
($N$(1440)1/2$^+$) and the $N$(1535)1/2$^-$, the $\Lambda$(1405) - $N^*$(1535) mass ordering, and the mass between the $\Lambda$(1405) and its spin-orbit partner, $\Lambda$(1520), are difficult to understand within a CQM framework.  
Models which include meson-baryon
interactions as further degrees of freedom have had success in describing such states~\cite{dalitz67, siegel88, kaiser97, recio04, lutz04}.
The $\Lambda$(1405) for example, appears to be dynamically generated from meson-baryon interactions to some extent~\cite{nacher03},
which is also confirmed by recent LQCD calculations~\cite{hall15}.
Models including vector meson-baryon interactions have predicted further dynamically generated states, for example $J^P = 1/2^-$ and $3/2^-$ around 2~GeV~\cite{gonzalez09, sarker10, oset10},
which may have been observed in $K^0\Sigma^+$ photoproduction~\cite{ralf, ewald14}.

Models using unquenched 5-quark components in baryons can also succesfully describe excitation spectra~\cite{zou08, wu10, wu09}.
Such models predict new, as yet unobserved $Y^*$ states, for example, a $\Sigma^*$ with $J^P = 1/2^-$ at a mass of 1380~MeV.

To understand the relevant degrees of freedom, a detector system with good acceptance at forward angles, and detection of mixed charge final states is crucial.  BGO-OD is ideally suited to these requirements.

\section{The BGO-OD experiment}

BGO-OD~\cite{bantes14, schmieden10, schmieden09} is a fixed target experiment at the ELSA electron accelerator facility~\cite{hillert06}.
An electron beam up to 3.5~GeV is incident upon a thin metal radiator
to produce energy tagged bremsstrahlung photons.
Linear and circular beam polarisations are both available (see ref.~\cite{Thomas}).

BGO-OD is split into two distinct parts: A central calorimeter region ($\theta = 25-155^0$) and a forward spectrometer ($\theta < 12^0$).
The acceptance hole between these is covered by a plastic scintillating detector (SciRi), and an MRPC (Daisy).  
Ref.~\cite{Thomas} shows a schematic of the current setup.

\begin{figure}[tbh]
\begin{center}
\includegraphics[width=1\textwidth]{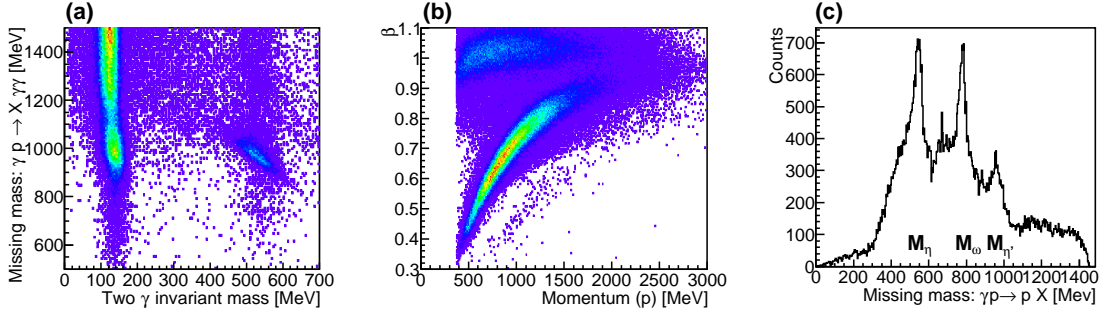}
\end{center}
 \caption{(a) The missing mass, $m_X$, from two photons detected in BGO ball versus the invariant mass of the two photons.  Peaks correspond to $p\pi^0$ and $p\eta$ final states.
 (b)  $\beta$ versus the measured momentum of charged particles in the forward spectrometer.  Prominent loci of charged pions and protons are visible.
 (c)  The missing (meson) mass from protons detected in the forward spectrometer.  Meson masses labelled inset.  (All analysis is preliminary, colour online.)}
 \label{f1}
 \end{figure}

A liquid hydrogen, deuterium or nuclear target is at the centre of the central region.
An MWPC surrounds the target for charged particle track reconstruction and accurate determination of the reaction vertex.  
Outside of this is a segmented cylinder of plastic scintillator material for charged particle identification via $\Delta E-E$.  
Surrounding this is the BGO ball; a segmented calorimeter of 480 bismuth germanate crystals.
The BGO ball is ideal for the reconstruction of photon four-momenta via clustering algorithms to characterise the electromagnetic showers in the crystals.
The separate time readout per crystal, with a resolution of approximately 3~ns, enables clean identification of neutral meson decays and reaction reconstruction (fig.~\ref{f1}(a)).

The forward spectrometer uses two scintillating fibre detectors (SciFi2 and MOMO) to track charged particles from the reaction or decay vertex.
Particles then proceed through the open dipole magnet, which operates at a maximum field strength of 0.45~T ($\int B dl \approx 0.45$~Tm).
Downstream are eight double layered drift chambers to track particle trajectories after curvature in the magnetic field.
Tracking algorithms are used to reconstruct the trajectory and particle momentum, $p$, with a resolution of approximately $0.015p$.
Time Of Flight (ToF) walls downstream from the drift chambers enable particle identification via the combination of momentum and $\beta$ (fig.~\ref{f1}(b,c)).

\section{Strangeness photoproduction proposals}

Approved  physics programmes include studies of pseudoscalar and vector mesons, in-medium modification of mesons, and reactions with open strangeness in the final state.   
The remainder of this paper focusses on the strangeness proposals, and preliminary analysis from the first data taking period.
BGO-OD is ideal for the investigation of reactions dominated by $t$-channel mechanisms due to the acceptance at forward angles with high momentum resolution.
Neutral particle identification in the central region aids the identification of final states of mixed charge.  This unique combination is suited
to the investigation of reaction channels with open strangeness and complicated final states. 
The first phase in this programme is the study of $K^0$ and $K^+$ reaction channels, which are described below.

Differential cross section measurements for $\gamma p \rightarrow K^0 \Sigma^+$ from the CBELSA/TAPS collaboration~\cite{ralf} exhibited a cusp like structure close to the 
$K^{*+}\Lambda$ and $K^{*0}\Sigma^0$ thresholds,
where the cross section reduced by a factor of four at forward angles.  
It was speculated that this may be due to subthreshold production of $K^*$ rescattering to $\pi^0$ or $K^0$~\cite{ewald14}.  BGO-OD will measure the beam asymmetry, $\Sigma$, over the $K^*$ threshold region, with high statistics by including data of both neutral and charged decays of the $K^0$ and $\Sigma^+$.

A paucity of data at centre of mass angles $\theta_{cm} < 15^0$ for $K^+Y$, prevent the constraining of isobar models~\cite{Bydzovsky12}.
The acceptance of the BGO-OD forward spectrometer is ideal to measure data at this important kinematic range. 
Data will also be taken over the disputed peak structure at $W = 1900$~MeV~\cite{glander04, tran98, bradford06, dey10, mccracket10}.

The second phase of proposals includes studies of $\gamma n \rightarrow K^0 Y$.
This is complementary to understand $K^+Y$ reaction mechanisms; dominant $t$-channel contributions are suppressed, and hadronic coupling constants used to describe the reaction mechanisms are related via SU(3) symmetry.
$K^+$ measurements will also be extended to search and establish higher $Y^*$ states.

\section{Preliminary analysis}
\subsection{$K^+Y$ identification in the BGO ball}

$K^+$ are stopped within the crystals of the BGO calorimeter up to a kinetic energy of approximately 400~MeV.  Using conventional clustering algorithms,
the measured energy signal is not proportional to the $K^+$ energy due to the subsequent weak decay within
the crystals.  A technique was developed with the Crystal Ball detector~\cite{jude14},
now implemented with the BGO ball, to use the time delayed weak decay as a means of $K^+$ identification
and energy reconstruction.

\begin{figure}[tbh]
\begin{center}
\includegraphics[width=0.3\textwidth]{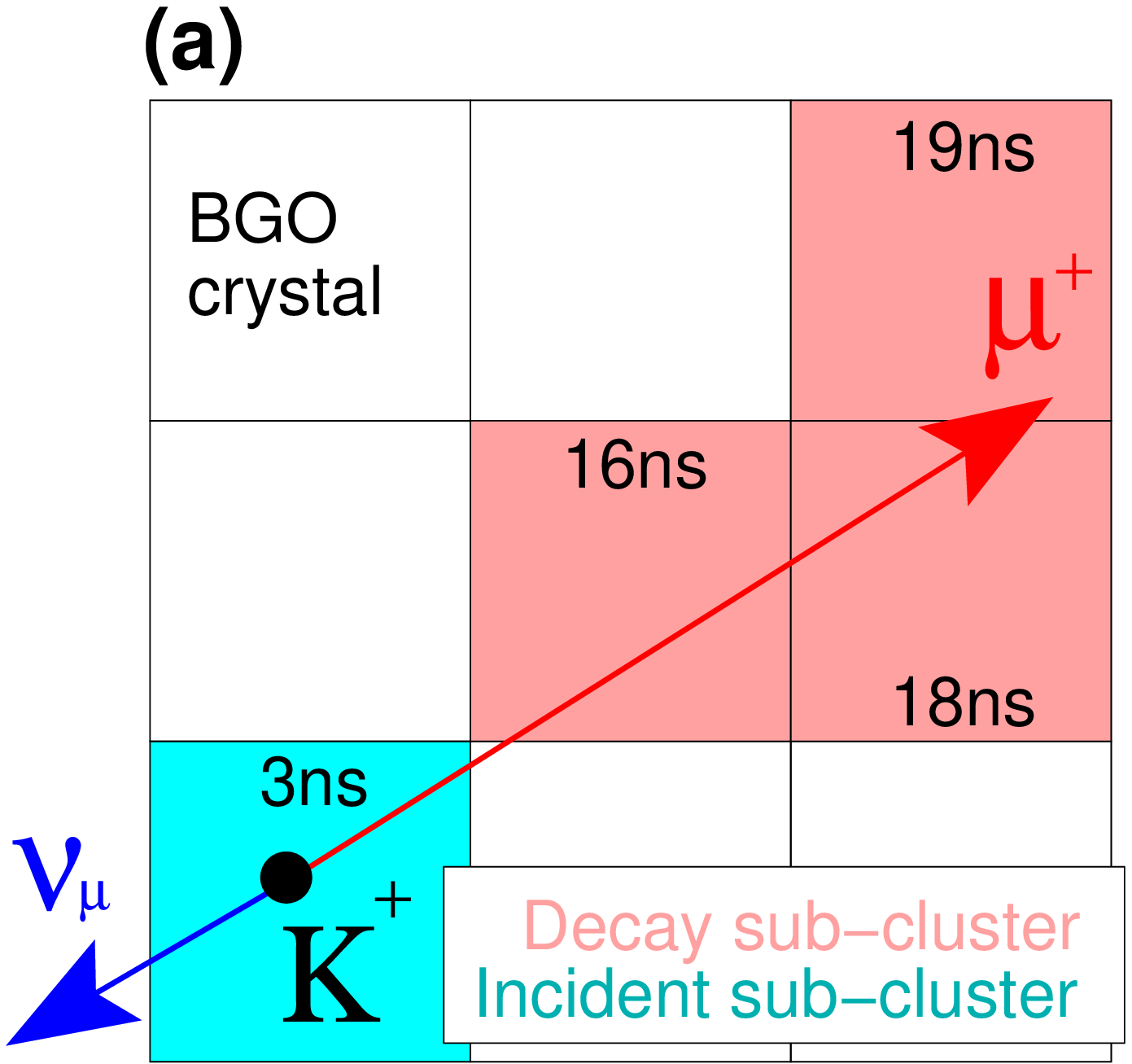}\includegraphics[width=0.7\textwidth]{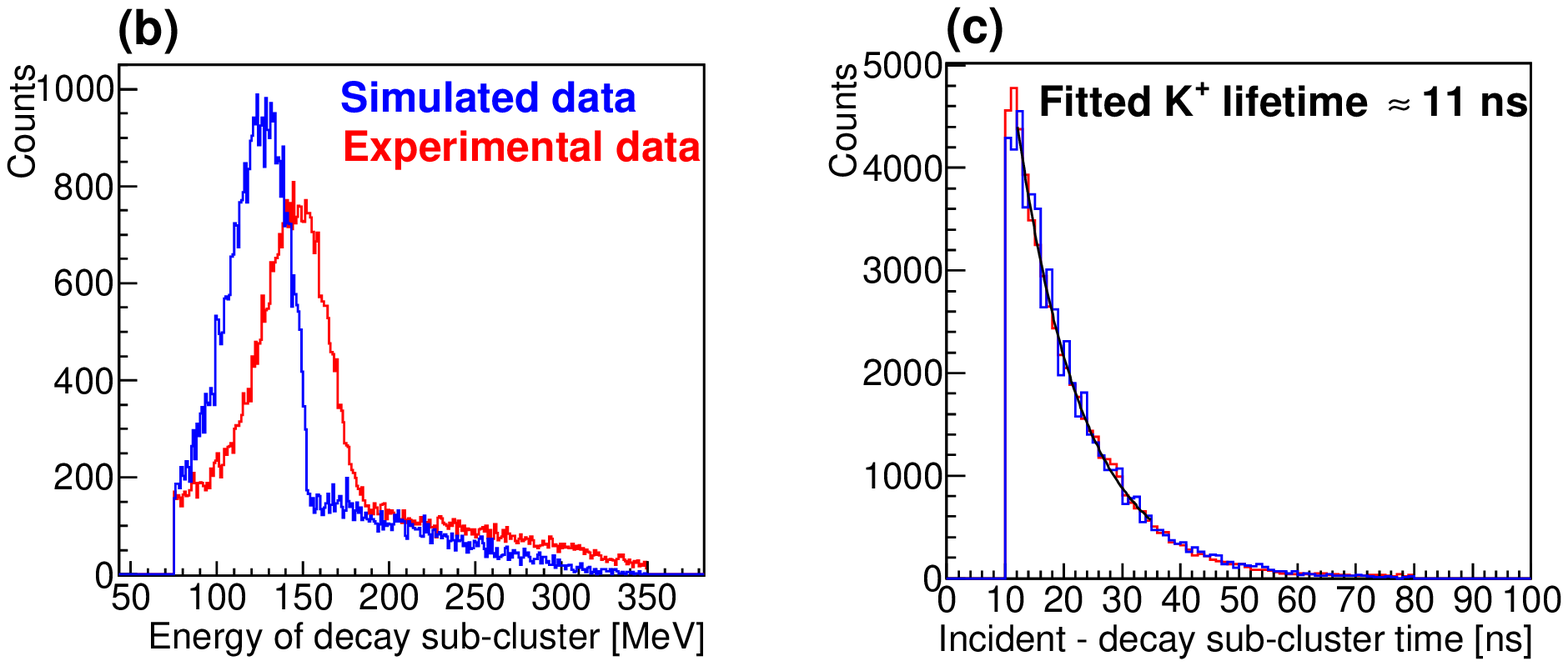}
\end{center}
\caption{$K^+$ identification in the BGO ball (preliminary). (a) An initial cluster is separated by time to an incident sub-cluster of crystals (blue) and decay sub-cluster (red).
The time inset each crystal indicates the time of the measured energy deposition.  
(b)  The energy of the decay sub-cluster.  The peak at 153~MeV corresponds to the $\mu^+$ energy deposition when the $K^+$ decays via the $\mu^+\nu_\mu$ mode at rest.
The shoulder extending to 350~MeV is due to the $\pi^+\pi^0$ decay mode.  The discrepancy between simulated and experimental data is due to the light collection behaviour of the photomultiplier tube not yet
implemented in the simulation.  (c) The time difference between the incident sub-cluster from stopping the $K^+$, and the decay sub-cluster from the $K^+$ decay.  An exponential fit yields
a lifetime close to the accepted value of 12~ns.}
\label{fig:kaondecay}
\end{figure}

$K^+$ have a lifetime of approximately 12~ns, with dominant decay modes to $\mu^+\nu_{\mu}$ and $\pi^+\pi^0$.
When a stopped $K^+$ decays in the BGO ball, energy depositions are recorded in a cluster of adjacent crystals.
This is identified using conventional clustering algorithms intended for electromagnetic shower reconstruction from incident photons.
The time resolution per crystal however, allows this cluster to be separated into an incident sub-cluster from stopping the $K^+$,
and a decay sub-cluster from the subsequent decay.  Fig.~\ref{fig:kaondecay} shows a schematic of the technique
and characteristic energy and time spectra.  Without using this method at the BGO-OD experiment, $K^+$ were only identified in the forward spectrometer.
This technique enables $K^+$ identification in the BGO ball,
vastly increasing the acceptance of reaction channels with open strangeness and vector mesons which decay to $K^+$.

\begin{figure}[tbh]
\begin{center}
\includegraphics[width=1\textwidth]{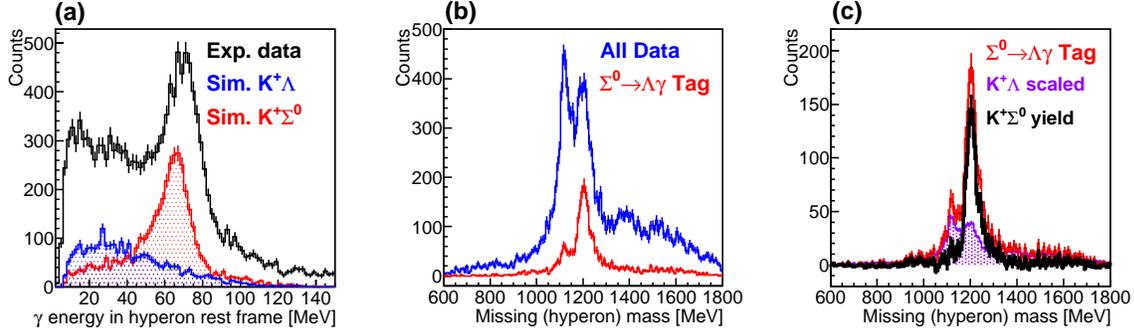}
\end{center}
\caption{(a) Photon energy after a Lorentz boost into the missing hyperon rest frame,
for experimental data (black line), simulated $K^+\Lambda$ (shaded blue line) and simulated $K^+\Sigma^0$ (shaded red line). 
(b) The missing hyperon mass from $K^+$ detected in the BGO for all experimental data (blue line)
and events where the $\Sigma^0 \rightarrow \Lambda \gamma$ is tagged (red line).
(c) Events where the $\Sigma^0 \rightarrow \Lambda \gamma$ is tagged (red line), the efficiency scaled background from misidentified $\Sigma^0 \rightarrow \Lambda \gamma$ candidates
(shaded purple line), and the yield of $K^+\Sigma^0$ after subtraction of background (thick black line). All preliminary.}
\label{fig:kaonmissmassbgo}
\end{figure}

A method to separate the final states, $K^+\Lambda$ and $K^+\Sigma^0$ was developed, incorporating this new $K^+$ identification technique.
A subset of events was selected with a $K^+$ and photon candidate and kept if the missing hyperon mass recoiling from the $K^+\gamma$ system was consistent with the $\Lambda$ mass.
The photon four momentum was subsequently boosted into the rest frame of the missing hyperon.  Fig.~\ref{fig:kaonmissmassbgo}(a) shows the energy of the boosted photon.
The peak corresponds to the $\Sigma^0 - \Lambda$ mass difference, which can be used to tag the $K^+\Sigma^0$ final state.
Using the simulated data shown in  fig.~\ref{fig:kaonmissmassbgo}(a), a selection over this peak yielded a detection efficiency of $\Sigma^0 \rightarrow \Lambda \gamma$ of approximately 60\%.  
A background resulting from the incorrect selection
of neutrons or photons from $\pi^0$ decays, which is evident in the simulated $K^+\Lambda$ events, was less than 10\%.
Fig~\ref{fig:kaonmissmassbgo}(b,c) show missing hyperon mass plots for $K^+$ detection in the BGO.  With the known efficiency
of the $\Sigma^0 \rightarrow \Lambda \gamma$ tag, the yields can be background subtracted to leave either $K^+\Lambda$ or $K^+\Sigma^0$ events (as shown in Fig~\ref{fig:kaonmissmassbgo}(c)).

\subsection{$K^+Y$ identification in the forward spectrometer}

\begin{figure}[tbh]
\begin{center}
\includegraphics[width=1\textwidth]{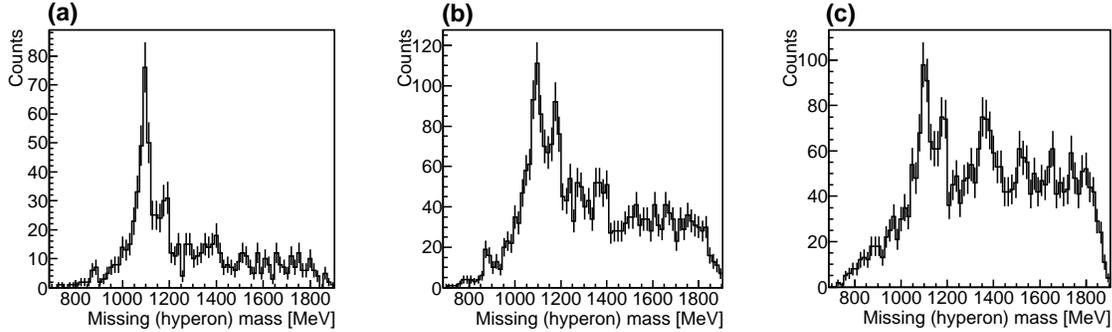}
\end{center}
\caption{The missing mass of the hyperon recoiling from the $K^+$ when identified in the forward spectrometer (preliminary).
(a) Requiring an energy deposition less than 250~MeV and a reconstructed $\pi^0$ in the BGO ball accentuates the $K^+\Lambda$ states.
(b) Requiring only an energy deposition less than 250~MeV accentuates the $K^+\Sigma^0$ and $K^+\Lambda$ final states.
(c) Requiring only a reconstructed $\pi^0$ in the BGO ball accentuates higher lying $Y^*$ states.}
\label{fig:kaonmissmass}
\end{figure}

Fig.~\ref{fig:kaonmissmass} shows the mass of the system recoiling from the $K^+$ when it is detected in the forward spectrometer.  
With the requirement of a $\pi^0$ reconstructed in the BGO,
$\Lambda$ and $\Sigma^0$ mass peaks are clear, and a spectrum of high lying states such as $\Sigma$(1385), $\Lambda$(1405), and $\Lambda$(1520) evident.

\section{Conclusions and acknowledgements}

Using commissioning data, the BGO-OD experiment and analysis is proven to be ready for studies in strangeness photoproduction.
Intense data taking periods for high statistics data has begun.

 We thank the technical staff of ELSA and the INFN, and the participating
 institutions for their invaluable contributions and acknowledge support from the Deutsche
 Forschungsgemeinschaft (SFB/TR16).


\begin{thebibliography}{9}


\bibitem{dalitz67} R.H. Dalitz, T.C. Wong and G. Rajasekaran, Phys. Rev. {\bf 153} (1967) 1617.
\bibitem{siegel88}  P.B. Siegel and W. Weise, Phys. Rev. {\bf C 38} (1988) 2221.
\bibitem{kaiser97}  N. Kaiser, T. Waas and W. Weise, Nucl. Phys. {\bf A 612} (1997) 297.
\bibitem{recio04}   C. Garcia-Recio, M.F.M. Lutz, and J. Nieves, Phys. Lett. {\bf B 582} (2004) 49.
\bibitem{lutz04}    M.F.M. Lutz and E.E. Kolomeitsev, Phys. Lett. {\bf B 585} (2004) 243.
\bibitem{nacher03} J.C. Nacher, E. Oset, H. Toki, A. Ramos, U.G. Meissner, Nucl. Phys. {\bf A 725} (2003) 181.
\bibitem{hall15}   J.M.M Hall {\it et al.}, Phys. Rev. Lett. {\bf 114} (2015) 132002.
\bibitem{gonzalez09} P. Gonzalez, E. Oset and J. Vijande, Phys. Rev. {\bf C 79} (2009) 025209.
\bibitem{sarker10} S. Sarkar {\it et al.}, Eur. Phys. J. {\bf A 44} (2010) 431.
\bibitem{oset10} E. Oset and A. Ramos, Eur. Phys. J. {\bf A 44} (2010) 445.
\bibitem{ralf} R. Ewald {\it et al.}, Phys. Lett. {\bf B 713} (2012) 180.
\bibitem{ewald14}  R. Ewald {\it et al.}, Phys. Lett. {\bf B 738} (2014) 268.
\bibitem{wu10}      Jia-Jun Wu, S. Dulat and B.S. Zou, Phys. Rev. {\bf C 81} (2010) 045210.
\bibitem{zou08}     B.S. Zou, Eur. Phys. J. {\bf A 35} (2008) 325.
\bibitem{wu09}      Jia-Jun Wu, S. Dulat and B.S. Zou, Phys. Rev. {\bf D 80} (2009) 017503.
\bibitem{bantes14}B. Bantes {\it et al.}, Int. J. Mod. Phys: Conf. Ser. {\bf 26} (2014) 1460093. 
\bibitem{schmieden10} H. Schmieden, Int. J. Mod. Phys. {\bf E 19} (2010) 1043.
\bibitem{schmieden09} H. Schmieden, Ch. Phys. {\bf C 33} (2009) 1146.
\bibitem{hillert06} W. Hillert, Eur. Phys. J. {\bf A 28} (2006) s01, 139. 
\bibitem{Thomas} T. Zimmermann, these conference proceedings.
\bibitem{Bydzovsky12}Bydzovsky and D. Skoupil, arXiv:1211.2684v1 (2012) Proceedings of SNP12.
\bibitem{glander04} K.H. Glander {\it et al.}, Eur. Phys. J. {\bf A 19} (2004) 251.
\bibitem{tran98} M.Q. Tran {\it et al.}, Phys. Lett. {\bf B 445} (1998) 20.
\bibitem{bradford06} R. Bradford and R.A. Schumacher {\it et al.}, Phys. Rev. {\bf C 73} (2006) 035202.
\bibitem{dey10} B. Dey, C.A. Meyer, M. Bellis, M.E. McCracken, M. Williams {\it et al.}, Phys. Rev. {\bf C 82} (2010) 025202.
\bibitem{mccracket10} M.E. McCracken, M. Bellis, C.A. Meyer, M. Williams {\it et al.}, Phys. Rev. {\bf C 81} (2010) 025201.
\bibitem{jude14} T.C. Jude, D.I. Glazier, D.P. Watts {\it et al.}, Phys. Lett. {\bf B 735} (2014) 112.

\end{thebibliography}
\end{document}